# CRISPR/Cas9 For Photoactivated Localization Microscopy (PALM)

Yina Zhu[1, #], Pingchuan Li[2, #], Paolo Beuzer[2, #], Zhisong Tong[2], Robin Watters[2], Dan Lv[3,4], Cornelis Murre[1], and Hu Cang[2, *]

[1] Department of Biology, University of California, San Diego, La Jolla, CA 92093, USA

[2] Waitt Advanced Biophotonics Center, Salk Institute for Biological Studies, 10010 North Torrey Pines Road, San Diego, CA 92037, USA

[3] School of Medicine, Nankai University, 94 Weijin Road, Tianjin 300071, China

[4] Department of Reproductive Medicine, Division of Gynecologic Oncology, Moores Cancer Center, University of California, San Diego, 3855 Health Sciences Drive, La Jolla, CA 92093, USA

# These authors contribute equally to the work.

* To whom correspondence should be addressed. hucang@salk.edu

### Abstract

We demonstrate that endonuclease deficient Clustered Regularly Interspaced Short Palindromic Repeats CRISPR-associated Cas9 protein (dCas9) fused to the photo-convertible fluorescence protein monomeric mEos3.1 (dCas9-mEos3.1) can be used to resolve sub-diffraction limited features of repetitive gene elements, thus providing a new route to investigate the high-order chromatin organization at these sites.

The development of Clustered Regularly Interspaced Short Palindromic Repeats CRISPR-associated enzyme Cas9 system has transformed the field of gene editing[1-11]. Since only a short small guide RNA (sgRNA) is required to target specific DNA loci, CRISPR/Cas9 is simpler to manipulate than previous target-specific gene editing tools including zinc finger (ZFN)[12] and transcription activator-like effector nucleases (TALEN)[13, 14]. Indeed, both ZFN and TALEN require custom-designed proteins for each DNA target, which is complex and time consuming. The ease of use of CRISPR/Cas9 has significantly lower the threshold of targeted genome engineering.

Recently, the application of CRISPR/Cas9 has been broadened to beyond gene editing[1-11]. Chen *et al* have demonstrated using CRISPR/Cas9 to guide fluorescence protein (GFP) to target gene loci including telomere and *MUC1-4*[1]. These results open new avenues for studying high-order chromatin structure[1, 15], which is a key contributor to epigenetic regulation by affecting the accessibility and affinity of regulatory proteins to their target sites. However, this study has only been carried out at low-, diffraction-limited resolution. Recently developed super-resolution imaging techniques, such as photoactivated localization microscopy (PALM)[16], fluorescence PALM (FPALM)[17] and stochastic optical reconstruction microscopy (STORM)[18, 19], are revolutionizing the study of cellular ultrastructure by providing unprecedented spatial resolutions[20, 21]. These super-resolution approaches demonstrate potentials to directly visualize chromatin and nuclear organization at the resolution of single-molecule level in single cells. The key to these approaches is a robust probe that can pinpoint a locus efficiently. Currently fluorescence in situ hybridization (FISH) has been the main tool for its high specificity[20]. However, the FISH procedure denatures DNA, which can induce artifacts for super-resolution microscopy[22]. Here, we demonstrate, for the first time, that by fusing endonuclease deficient dCas9 to the photo-convertible fluorescence protein mEos3.1, we can use dCas9-mEos3.1 for PALM to resolve sub-diffraction limited features of repetitive DNA sequences in telomeres, paving the way for high resolution studies of chromatin organization and epigenetic regulation at these loci.

We chose green-to-red photoconvertible protein mEos3.1[23], because it is a true monomeric derivative of mEos2. As one of the best overall performing photoactivatable fluorescence proteins for (F)PALM, mEos3.1preserves the similar level of high photon output of mEos2, critical for super-resolution in PALM[24], while overcomes the dimerization problem of mEos2 at high concentrations. We fused mEos3.1 to the c-terminal of dCas9 (D10A). To ensure the fused protein entering the nucleus, two copies of nuclear localization signal (NLS) sequences were placed before and after dCas9[1, 2, 9, 10]. To minimize free diffusing dCas9-mEos3.1, which contributes to the background noise, we used a Tet-Off tetracycline-controlled transcriptional activation system to control the expression level of dCas9-mEos3.1. We generated stable U2OS (human osteosarcoma) and MEF (mouse embryonic fibroblast) cell lines expressing Tet-Off dCas9-mEos3.1by lentivirus transduction (Fig. 1). The cells were cultured in medium with a doxycycline (Dox) concentration of 100ng/mL to suppress the expression of excessive dCas9-mEos3.1for at least 48 hours before PALM experiments. To demonstrate the super-resolution capability of CRISP/Cas9, we imaged the telomeres in both U2OS and MEF cells. Telomeres are repetitive nucleotide sequences at each end of chromatids. Telomeric repetitive sequences have previously been shown to represent a good target for dCas9-EGFP[1]. We used the previously described sgRNA, containing a 22 nt telomere targeting sequence, with the murine polymerase III U6 promoter. Lentivirus has been used to transduce sgRNA into stably expressing dCas9-mEos3.1 U2OS or dCas9-mEos3.1 MEF cells. At the time of the transduction, doxycycline was removed from the medium to induce the expression of dCas9-mEos3.1. 48hr after the transduction, the cells were fixed for super-resolution imaging. To further reduce the background from free dCas9-mEos3.1, samples were stained in cytoskeleton (CSK) buffer on ice for 4mins prior to fixation.

To confirm dCas9-mEos3.1 specifically targeting to telomeres, we labeled the telomere nucleoprotein complex Telomeric Repeat-binding Factor 2 (TRF2) by immunofluorescence and performed dual-color confocal fluorescence imaging in U2OS cells. dCas9-mEos3.1 co-localize with TRF2, demonstrating the specificity and efficiency of dCas9-mEos3.1 as telomere label (Fig. 1).

One of the critical steps in imaging telomeres by using PALM is to properly set the focus of the microscope. Typically, a 405nm laser is needed to photoconvert mEos3.1 from green to red, and mEos3.1 fluorescence is collected through a red fluorescence protein (RFP) filter set. However, telomeres are not visible at first in the RFP channel before the convert. Thus, we use the green fluorescence of dCas9-mEos3.1 to set the focus coarsely, and then use the 405nm laser to convert a small portion of dCas9-mEos3.1 to red and determine the focus precisely in the RFP channel. The focus is then locked by a focus-tracking system (ASI Imaging) for PALM. PALM reveals sub-diffraction details of the telomere structures not detectable by conventional fluorescence microscopy in the MEF cells (Fig 2a-c). We have analyzed the distribution of the size of telomeres (Fig. 2c) and calculated the width of single telomere as twice of the standard deviations of the observed single molecule events at each telomere. The analyzed 91 telomeres from 8 cells show a broad distribution in size. The average is 179 ± 49nm (Fig. 2c), which is consistent with a recent study on telomere by using STORM[20] with peptide-nucleic acid (PNA) probes. We then imaged telomeres in U2OS cells (Fig. 3). U2OS cell line belongs to 10-15% of cancer cell lines that use the homologous recombination based ALT (alternative lengthening of telomeres pathway) mechanism to maintain the length of telomere for unlimited cellular proliferation[25, 26]. Telomeres of ALT-positive cells are heterogeneous in size and repeat length. While the telomeres in U2OS appeared to be similar in size of MEF cells, in a conventional epi-fluorescence image with diffraction-limited resolution (Fig. 3a), PALM shows that the telomeres of U2OS are slightly larger (Fig. 3b). The average diameter of 41 telomeres observed from 4 cells is 215 ± 88 nm (Fig. 3c). This suggests that some of the analyzed telomeric foci may contain fusion of multiple telomeres[25].

We tested another endonuclease deficient mutant dCas9 (D10A and H840A). To control the expression level we engineered a Tet-On inducible dCas9-mEos3.1, and used lentivirus to generate a stable expressing MEF cell line. We found that the basal level expression of mutant dCas9(D10A and H840A)-mEos3.1was sufficient for single-molecule imaging. The same sgRNA-telomere was tested. PALM experiments show essentially the same results (Supplementary Figure).

In summary, we have demonstrated that the powerful locus-targeting dCas9 system can be exploited for single molecule super-resolution microscopy PALM. Our study undercover the difference in telomere size in ALT-positive U2 OS and negative MEF cells, which is difficult to quantify with a conventional microscope. We envision that several lines of improvements of CRISPR/Cas9 that are currently undergoing will accelerate its use for studying genome structure in near future. For example, the development of orthogonal sgRNA[27] will enable multi-color dCas9 probes for multi-color PALM; the improved high specificity design of sgRNA[28] will reduce off-target rate of dCas9 and increase the image contrast.

## Method

### Plasmid Preparation

Tet-Off-dCas9(D10A)-mEos3.1. In order to fuse dCas9(D10A) with mEos3.1, we first cloned dCas9 into pCMV-mEos3.1 N1 (a gift from Jennifer Lippincott-Schwartz, NIH). Px335 (D10A, Addgene, 44335) was used as template for PCR. A BamHI site was introduced at 3' terminal of dCas9, and then mEos3.1 was fused into Px335 between BamHI and EcoRI sites. The fusion gene, AgeI-NLS-dcas9-NLS-mEos3_EcoRI was subcloned into a lenti-Tet-Off vector pBoB-TA1-R2-corrected 120803 from Verma lab, Salk Institute. This vector was then called Tet-Off-dCas9(D10A)-mEos3.1.

Tet-On-dCas9(D10A, H840A)-mEos3.1. We first cloned the double mutant dCas9 into pCMV-mEos3-N1. Two NLS are placed before and after the dCas9. pdCas9-humanized(D10A H840A, Addgene, 44246) was used as template for PCR. The primers used are: F-BamHI-NLS-dcas9: 5-AAAGGATCCGCCACCATGAGCCCCAAGAAGAAGAGAAAGGTGGAGGCCAGCGACAAGAAGTATTCTATC-3, and R-dcas9-NLS-AgeI: 5-AAAACCGGTCCTACCTTTCTCTTCTTTTTGGATCAGCTCCCTCATCCCC-3. The resulting fusion gene, BamH1-NLS-dcas9-NLS-Age1 was subcloned into pMIR40, a lenti-Tet-On vector from Verma lab, Salk Institute.

Telomere sgRNA were prepared following the previously reported procedure, with pgRNA-humanized (Addgene, 44248) as template. The primers used are: PCR-L-primer: 5-GGAGAACCACCTTGTTGGTTAGGGTTAGGGTTAGGGTTAGTTTAAGAGCTATGCTGGAAACAGCATAGCAAGTTTAAATAAGGCTAGTCCGTTATCAAC-3. The telomere target sequence was represented by an underline. PCR-R-primer: 5-CTAGTACTCGAGAAAAAAAGCACCGACTCGGTGCCAC-3. For PALM, we deleted RFP mCherry gene from the above sgRNA-Tel vector via AgeI/XmaI double digestion followed with re-ligation. All plasmids were confirmed by Sanger sequencing (Retrogen).

### Cell Culture, Lentiviral Production, and Cell Transduction

U2OS, and MEF cell lines were maintained in DMEM (10% FBS, 1% antibiotic) at 37° and 5% CO2 in humidified incubator. The Lenti-X 293T Cell Line was ordered from Clontech. Lentivirus was prepared following the recommended protocol from the company. Lenti-X Concentrator (Clontech) was used to concentrate virus. Virus was produced by transient calcium phosphate mediated transfection of viral pMD2.G, psPAX, and Tet-Off-dCas9-mEos3.1, or Tet-On-dCas9-mEos3.1, or sgRNA into 293T. Virus was harvested 48 hr post transfection.

To generate a stable cell line expressing Tet-Off-dCas9-mEos3.1, we cultured this cell line with 100ng/ml Dox to prevent the expression of excess dCas9 protein. We then transduced the dCas9-mEos3.1stable expressing cell line with telomere sgRNA lentivirus. We performed lentiviral spin-transduction of telomere sgRNA at 1,600 rpm with 10μg/ml polybrene. Right before the transduction, we changed cell medium with a Dox free medium to allow the expression of dCas9 protein. The Tet-On-dCas9-mEos3.1cell lines were used without Dox. The basal level expression of dCas9-mEos3.1is sufficient for imaging. We fixed cells for PALM imaging 48 hours after the transduction of sgRNA.

**Immunofluorescence**

U2OS and MEF were grown on cover slips and pre-extracted with CSK buffer (0.1% Triton X-100, 20 mM Hepes-KOH pH7.9, 50 mM NaCl, 3 mM MgCl2, and 300 mM Sucrose) on ice for 4 min. Cells were then fixed in 4% para-formaldehyde in PBS for 10 min; washed with PBS, blocked and permeabilized with 2% (w/v) bovine serum albumin (BSA) and 0.25% (v/v) Triton X-100 in PBS for 30 min. Cells were then incubated at 37° for 30 min with rabbit anti-TRF2 primary antibodies (a gift from Karlseder lab, Salk Institute) diluted to 1.25μg/mL in blocking buffer, washed with PBS; and stained for 40 min with Alexa647 donkey anti-rabbit secondary antibodies (Life Tech) diluted to ~4 μg/mL in blocking buffer. The coverslips were then mounted on a 1mm slide with mounting medium (Vecta-shield). Dual-color fluorescence confocal-images of TRF2 immunofluorescence and telomere-bound dCas9-mEos3.1were taken by a Zeiss LSM710 confocal microscope (Waitt Advanced Biophotonics Center, the Salk Institute). ImageJ was used to analyze the images.

**PALM Microscopy**

Coverslips with MEF and U2OS samples were glued and sealed on the bottom of a petri-dish with a 0.8 cm hole in the center. The dish was filled with PBS phosphate buffered saline. PALM was carried out on a Nikon Ti-U inverted microscope with a 1.49NA 60x oil-immersion objective lens. A 140mW 561nm laser (Cobalt Jive) was used for excitation, a 20mW 405nm laser (Phoxx) was used to activate mEos3.1. The samples were first imaged by a conventional epi-fluorescence with a GFP filter set (Chroma) to determine the focus; the microscope is then switched to PALM. A 406/561 dual line dichroic mirror (Chroma) was used to couple the laser beams into the samples. A 561nm notch filter (Semrock), and a 488nm long-pass filter (Thorlabs) were used to remove the excitation and activation laser beams. No extra emission filter was used. A custom control system was used to alternate between the 405nm activation and the 561nm excitation laser for PALM. A focus locking system (ASI) was coupled with a piezo-stage (Physik Instrument) to avoid the focus drift. The PALM images were reconstructed from movies of 2,000 – 6,000 frames, recorded at 25Hz. An inclined incidence illumination geometry (HILO) was used to reduce background fluorescence.

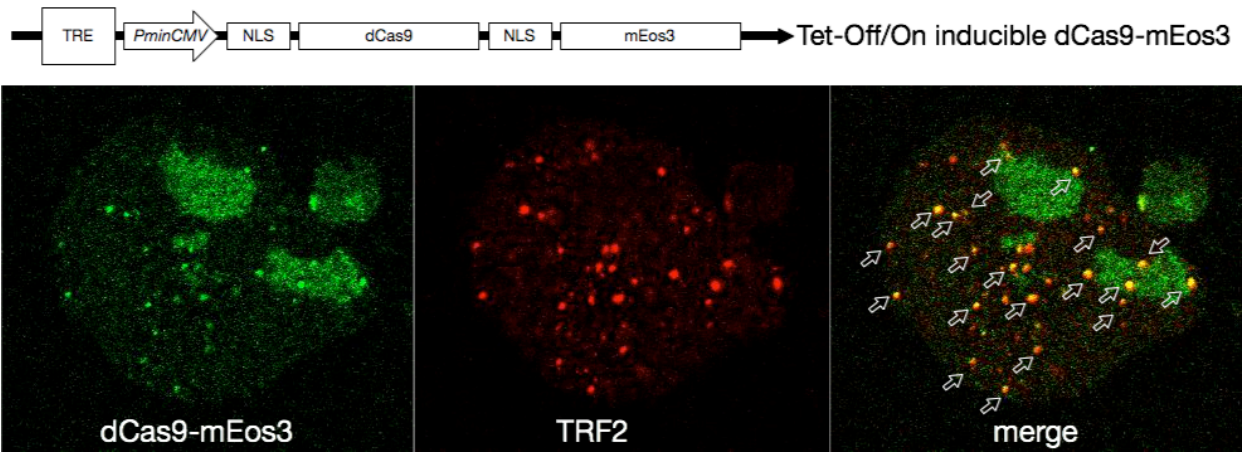

**Fig. 1 Colocalization of dCas9-mEos3.1(green) with TRF2 (red) at telomeres in U2OS cells.**
A Tet-Off inducible system is used to control the expression of dCas9-mEos3.1. Two copies of nuclear localization signal (NLS) were placed before and after dCas9 to ensure the fused dCas9-mEos3.1entering nucleus. dCas9-mEos3.1and immunofluorescence of TRF2 in U2OS cells are shown in the left and middle panel, respectively. Prior to the fixation, soluble proteins were removed with 0.1% CSK Triton X-100 buffer. Left panel: dCas9-mEos3.1, the fluorescence from mEos3.1is taken with a GFP filter set. Middle panel: immunofluorescence of TRF2 labeled with Alexa647. Right panel: merged image. The images were taken with a Zeiss LSM-700 confocal microscope. Colocalization sites (yellow) are highlighted by the arrows in the merged figure.

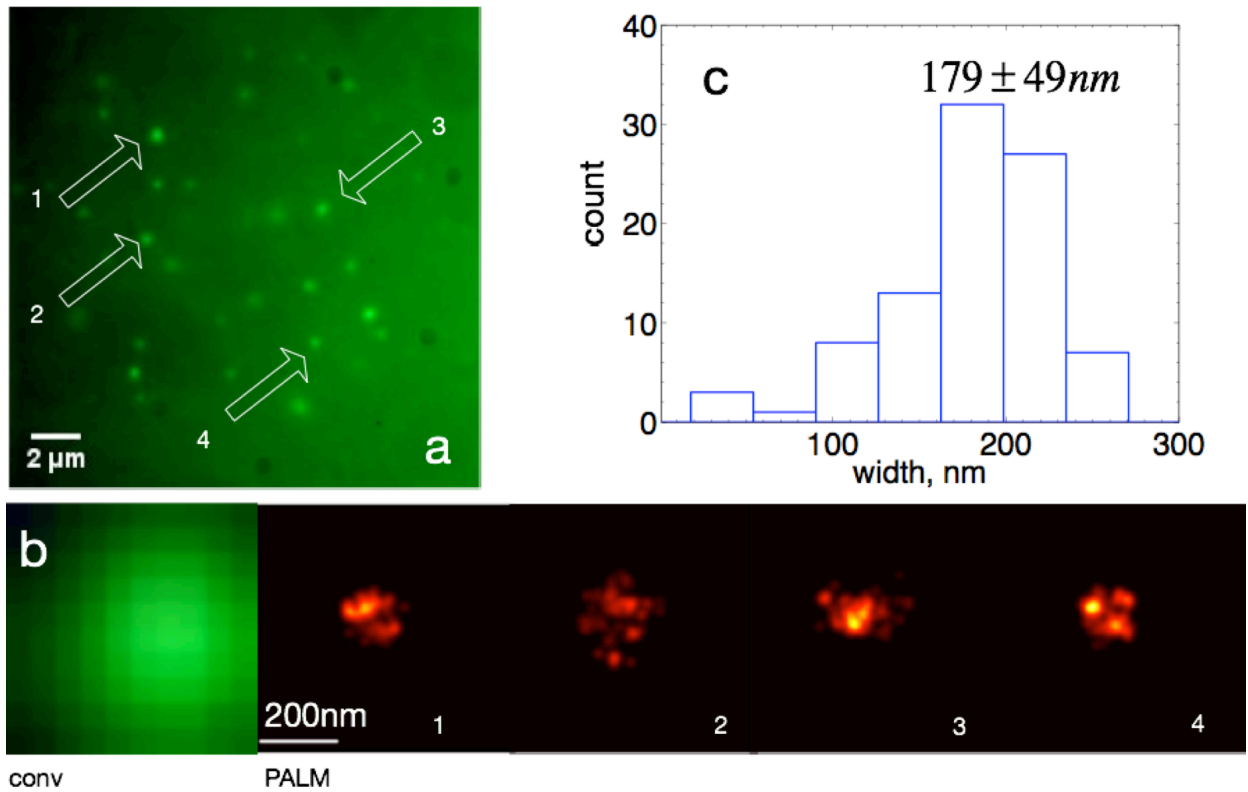

**Fig. 2 PALM image of telomeres in MEF cells.**

**(a).** A typical diffraction-limited epi-fluorescence image of dCas9-mEos in a MEF cell. The image was taken with a GFP filter set. The bright dots correspond to telomeres. A zoomed in image of telomere marked with number 1, is shown in **(b)**. The corresponding PALM images of the ones marked with arrows are also shown in (b). The reconstructed PALM images reveal details that are not resolved in the conventional epi-fluorescence image. The scale bar in panel B is 200nm. **(c)** PALM shows a broad distribution of the widths. The mean width of the telomeres, measured by the standard deviation, is $179 \pm 49$ nm.

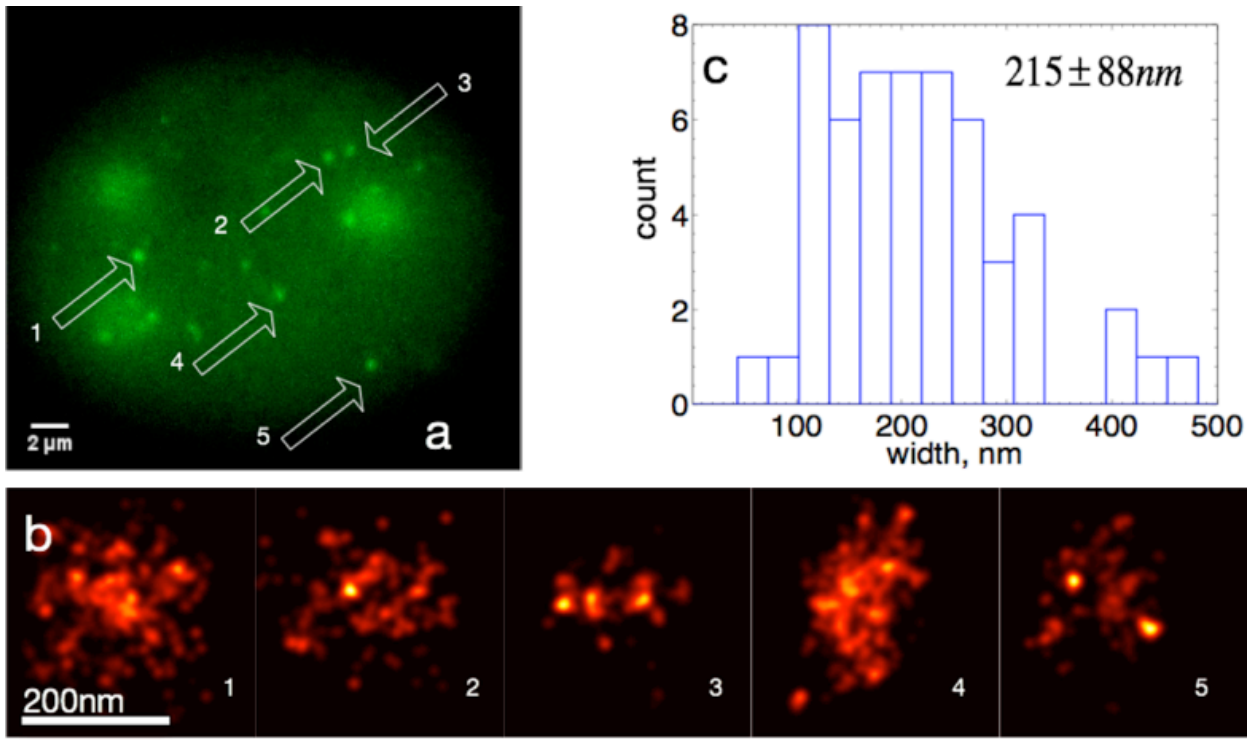

**Fig. 3 PALM image of telomeres in U2OS cells.**

**(a).** A typical diffraction-limited epi-fluorescence image of dCas9-mEos in a U2OS cell. A GFP filter set was used. **(b)** Reconstructed PALM images of the telomeres marked with arrows. Telomeres in U2OS are larger than in MEF cells. The scale bar is 200nm. **(c)** The mean width of the telomeres in U2OS cells, measured by the standard deviation, is $215 \pm 88$ nm.

12

**Supplementary Figure**

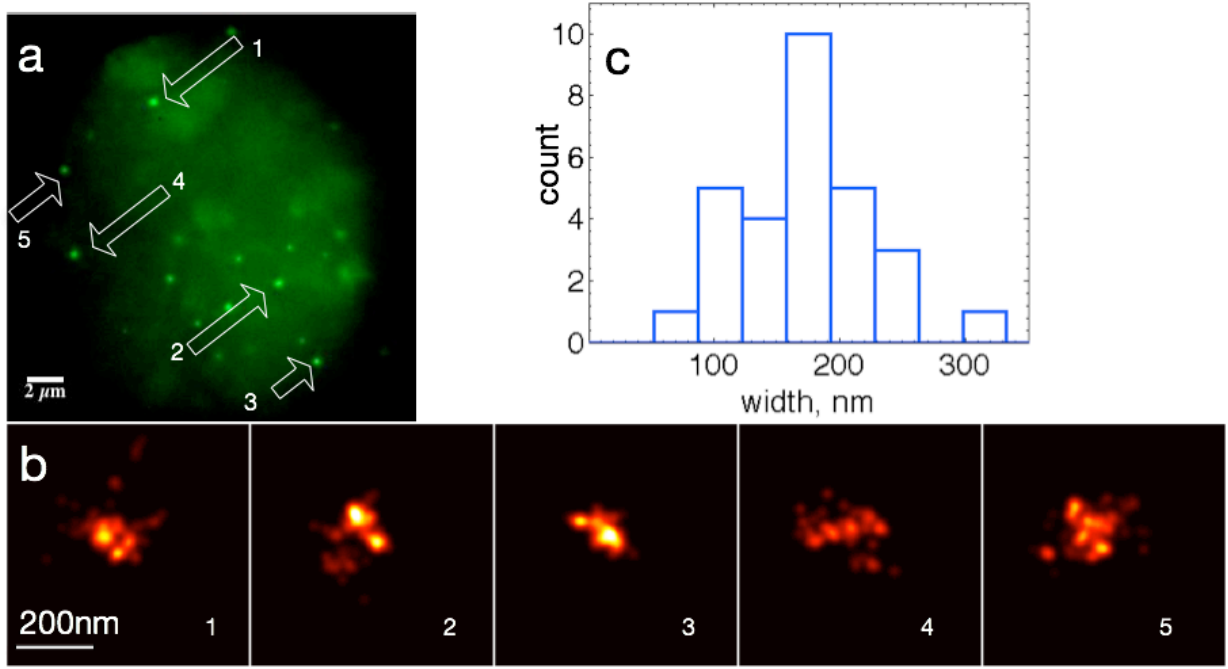

**PALM image of telomeres in MEF cells with Tet-On-dCas9(D10A, H840A)-mEos3.1.**

**(a).** A typical diffraction-limited epi-fluorescence image of dCas9(D10A, H840A)-mEos3.1 in a MEF cell. A GFP filter set was used. **(b)** Reconstructed PALM images of the telomeres marked with arrows. The scale bar is 200nm. **(c)** The mean width of the telomeres measured from 4 cells is $175 \pm 52$nm.

13